\documentclass[12pt,manuscript]{aastex}
\begin{document}

\title{Light Isotope Abundances in Solar Energetic Particles \\ measured by the
 Space Instrument NINA}

\author{A.  Bakaldin, A. Galper, S. Koldashov, M. Korotkov, A. Leonov, V. 
Mikhailov,
A. Murashov, S. Voronov}
\affil{Moscow Engineering Physics Institute, Kashirskoe Shosse 31, 115409 
Moscow, Russia}
\author{V. Bidoli,  M. Casolino, M. De
Pascale,  G. Furano, A. Iannucci, A. Morselli,  P. Picozza,  R.
Sparvoli\altaffilmark{1 }} \affil{Univ. of Rome ``Tor Vergata''
and INFN sezione di Roma2, Via della Ricerca Scientifica 1,
I-00133 Rome, Italy}
\author{M. Boezio, V. Bonvicini, R. Cirami, A. Vacchi, N. Zampa}
\affil{Univ. of Trieste and INFN sezione di Trieste, Via A.
Valerio 2, I-34147 Trieste, Italy}
\author{M. Ambriola, R.
Bellotti, F. Cafagna, F. Ciacio, M. Circella, C. De Marzo}
\affil{Univ. of Bari and INFN sezione di Bari, Via Amendola 173,
I-70126 Bari, Italy}
\author{O.  Adriani, P.
Papini, S. Piccardi, P. Spillantini, S. Straulino, E. Vannuccini}
\affil{Univ. of Firenze and INFN sezione di Firenze, Largo Enrico
Fermi 2, I-50125 Firenze, Italy}
\author{S. Bartalucci, M. Ricci}
\affil{INFN Laboratori Nazionali di Frascati,  Via Enrico Fermi 40,  I-00044 
Frascati, Italy}
\and
\author{G. Castellini}
\affil{Istituto di Ricerca Onde Elettromagnetiche CNR, Via
Panciatichi 64, I-50127 Firenze, Italy }

\altaffiltext{1}{Dept. of Physics, Univ. of Rome ``Tor Vergata''
and INFN - Via della Ricerca Scientifica 1,  00133 Rome, Italy.
Email: Roberta.Sparvoli@roma2.infn.it. Tel: +39-06-72594575 - Fax:
+39-06-72594647}

\begin{abstract}

This article reports nine  Solar Energetic Particle events
 detected by the instrument NINA between
October 1998 and April 1999. NINA  is a silicon-based particle
detector mounted on-board the Russian satellite Resurs-01-N4,
which has flown at
 an altitude of about 800 km in polar
inclination since July 1998.

For every solar event the power-law $^4$He spectrum across 
the energy interval 10--50 MeV n${-1}$ was reconstructed,
and spectral indexes, $\gamma$,   from 1.8 to 6.8  extracted. Data of
$^3$He and $^4$He were used to determine the $^3$He/$^4$He ratio,
that  for some SEP events indicated an enrichment  in $^3$He. For
the 1998 November 7 event the ratio reached a maximum value of
$0.33\pm 0.06$, with spectral indexes of $\gamma$ = 2.5 $\pm$ 0.6
and $\gamma$ = 3.7 $\pm$ 0.3 for $^3$He and $^4$He, respectively.
The $^3$He/$^4$He ratio averaged over the remaining events was
$0.011\pm 0.004$.

For all events   the deuterium-to-proton ratio was determined.
The average value over all events was (3.9$\pm$1.4)$\times$
10$^{-5}$ across the energy interval 9--12 MeV/n. For one event 
(1998 November 24) this ratio yielded approximately  10 times
higher than normal coronal values.

Upper limits on the $^3$H/$^1$H  counting ratio for all events
were determined. For the 1998 November 14 SEP event the high flux
of heavy particles detected  made it possible to reconstruct
 the carbon  and oxygen flux.

\end{abstract}


\section{Introduction}

In 1970 \citet{hsieh}  discovered Solar Energetic Particle (SEP)
events with a greatly enhanced abundance of the rare isotope
$^3$He, with the IMP-4 satellite. Following this, many other
experiments started to study solar $^3$He-rich events. In such
events the ratio between $^3$He and $^4$He  is strongly enhanced
with respect to solar abundances ($\sim 4\times 10^{-4}$)
measured in the Solar Wind \citep{Coplan1985,Bodmer1995}. It was
subsequently found that in these events abundances of heavy
elements were also unusual, with a  Fe/O ratio about 10 times the
value in the corona [see the review by \citet{Reames1999}].

 The first hypothesis  suggested that nuclear reactions of accelerated particles
in the ambient of the solar atmosphere are the most probable
source of $^3$He. The presence of such reactions was
 independently confirmed by the detection of the 2.2 MeV
neutron-proton recombination $\gamma$-line in solar flares
\citep{chupp1973}. Nowadays,  $\gamma$-ray spectroscopy
\citep{mand1999} shows that there is  presence of $^3$He at the
flare sites.

It is known that  an enormous abundance of $^3$He in SEPs does
not correspond to an overabundance of  $^2$H and $^3$H. While
solar $^3$He was detected by many observers, solar deuterium and
tritium have proven to be very rare and difficult to detect in
SEP events
\citep{anglin1975,mewaldt1983,vanholle1985,1986mcguire}. Measured
abundances of $^2$H and $^3$H in Solar Energetic Particles are
consistent with the {\it thin target model} of flares
\citep{ramaty1974}. However, it was shown \citep{anglin1975} that
this  model could not explain the 1000-fold enhancement in the
$^3$He/$^4$He ratio  and also the enrichment of heavy elements.
It became clear that selective mechanisms of $^3$He acceleration
were required \citep{fisk1978}.

Solar Energetic Particles are now believed to come from two
different sources. The SEPs from solar flares have a 1000-fold
enhancement in the $^3$He/$^4$He ratio and an enhanced number of
heavy ions with respect to solar abundances  because of resonant
wave-particle interactions at the flare site, where the ions are
highly stripped of orbital electrons by the hot environment.
However, the most intense SEP events, with particles of the
highest energies, are produced by accelerations at  shock waves
driven by Coronal Mass Ejections (CMEs). On average, these
particles directly reflect the abundances and temperature of
ambient, unheated coronal material. Various aspects of gradual
and impulsive SEP events have been compared in a variety of
review articles \citep[and references therein]{Reames1999}.

Flares and CMEs can occur separately. Most flares are not
accompanied by a CME, whereas many fast CMEs that produce gradual
SEP events  have associated flares. The most interesting events
are therefore the "pure" ones, where the two mechanisms are not
acting together and it is  easier therefore to distinguish and
characterize the acceleration processes involved.

The $^3$He/$^4$He ratio can be used to characterize the two types
of event. $^3$He/$^4$He $\sim$ 1 is typical in impulsive events,
while $^3$He/$^4$He $<$ 0.01 indicates gradual SEP events.
However, this ratio is difficult to measure and not available for
a large sample of events \citep{chen1995,mason1999}. Precise
measurements of the $^3$He/$^4$He ratio can provide new
constraints on existing theories that discuss $^3$He acceleration
mechanisms and the propagation processes.

\vspace{0.7cm}

The aim of this article is to present measurements of light
isotope abundances in  Solar Energetic Particle events detected
in the period October 1998 -- April 1999 by the instrument NINA.
 NINA was launched on 1998 July 10 from the Baikonur
launch facility, in Kazakhstan,   on board of the Russian
satellite Resurs-01-N4. The orbit of the spacecraft is
sun-synchronous, with an inclination of 98 degrees   and an
altitude of approximately 800 km. In \S 2 we describe the
instrument  briefly, in \S 3 we discuss the algorithms of event
selection and particle identification, and the background
analysis. Results are presented in \S 4,  and possible
interpretations are given in \S 5.

\section{The instrument}

The instrument NINA is optimized for the detection of galactic,
solar and anomalous cosmic rays in the energy window 10--200 MeV
n$^{-1}$. It   is   composed of 16  planes, each
 made of two n-type silicon detectors, 6$\times$6 cm$^2$, segmented in 16 strips
and orthogonally mounted so to provide the X and Y information of
the particle track. The thickness of the first two detectors is
150 $\mu$m; all the others  are 380 $\mu$m thick, for a total of
11.7 mm of active silicon.

The 16  planes are vertically stacked. The interplanar distance
is  14 mm, but the first and second planes are separated by 85 mm,
for a better measurement of the particle incident angle. The total
telescope height is 29.5 cm.
 An aluminum layer 300  $\mu$m thick
 covers the detector.
 More details about the detector configuration can be found in
\citet{nina1,nina2,nina3}.

A veto system,
 ensuring the containment of  particles entering the detector from above,
is implemented by setting in anticoincidence  strips 1 and 16 of
planes  2  to 15 (Lateral Anticoincidence),  and  all strips of
plane 16  (Bottom Anticoincidence). The request of containment
inside the telescope defines the upper energy limits for particle
detection,  which is equal to about 50 MeV n$^{-1}$ for $^4$He.

The  threshold for  energy deposits in the silicon layers is set
to 250 keV (Low Threshold) in order to eliminate all relativistic
protons which release only $\sim$ 100 keV in 380 $\mu$m of
silicon. A second threshold,  at 2500 keV (High Threshold),
further restricts  the detection of protons. All results presented
in this paper come from data acquired in High Threshold Mode.

\section{Data analysis}
The study of SEP rare isotope abundances  requires a careful
analysis of the data collected. The background includes either
noise  and unidentifiable events, which can be eliminated by an
off-line dedicated  selection algorithm, and particles produced by
secondary interactions in the material surrounding the  NINA
silicon tower. In order to detect only solar particles,  the
solar quiet background must be excluded also.

\begin{figure}
\epsscale{.9} \plotone{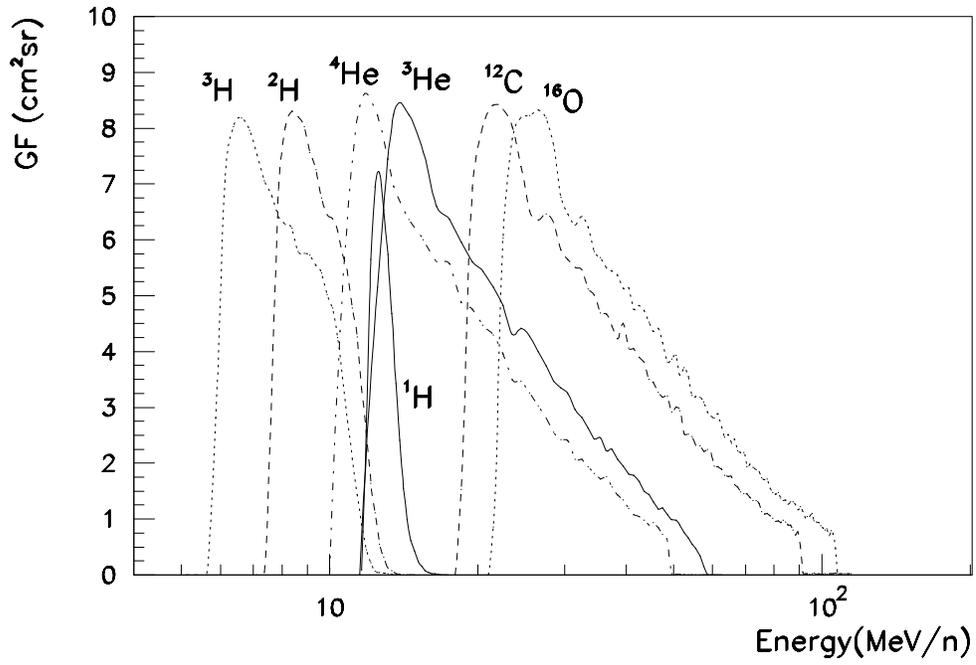}
   \caption{Geometric factor, GF, of NINA  for $^1$H, $^2$H, $^3$H, $^3$He, 
$^4$He, $^{12}$C,
   and $^{16}$O  in High Threshold Mode.}
\label{gf}
\end{figure}

Table \ref{thre} shows the energy window of NINA for the atomic
species mostly studied during SEP events, in High Threshold
configuration. Figure \ref{gf} shows the Geometric Factor, GF, for
the same isotopes.

\subsection{Event selection and mass reconstruction}
The optimal performance of NINA
 in terms of charge, mass and energy
determination is achieved by requiring the full containment of the
particle inside the detector with the Lateral and Bottom
Anticoincidence. Despite the presence of the Lateral
Anticoincidence, some particles  leave the detector between
silicon planes. In order to eliminate this effect, tracks with
energy deposits in strips 2 or 15 for any of the layers of the
silicon tower are   rejected off-line.
 This requirement  reduces the Geometric Factor of the instrument, and
 this is  taken into account in  calculations (Figure \ref{gf}).
The segmented nature of the  detector, in addition to measuring
the total energy released, allows a very precise determination of
the topology of the particle's path inside the instrument. With
this information it is possible to build a dedicated off-line
track selection algorithm  which rejects upward moving particles,
tracks accompanied by nuclear interactions, and  events
consisting of two or more tracks. A detailed description of the
selection algorithm can be found in \citet{nina3}.

\begin{figure}
\epsscale{.7} \plotone{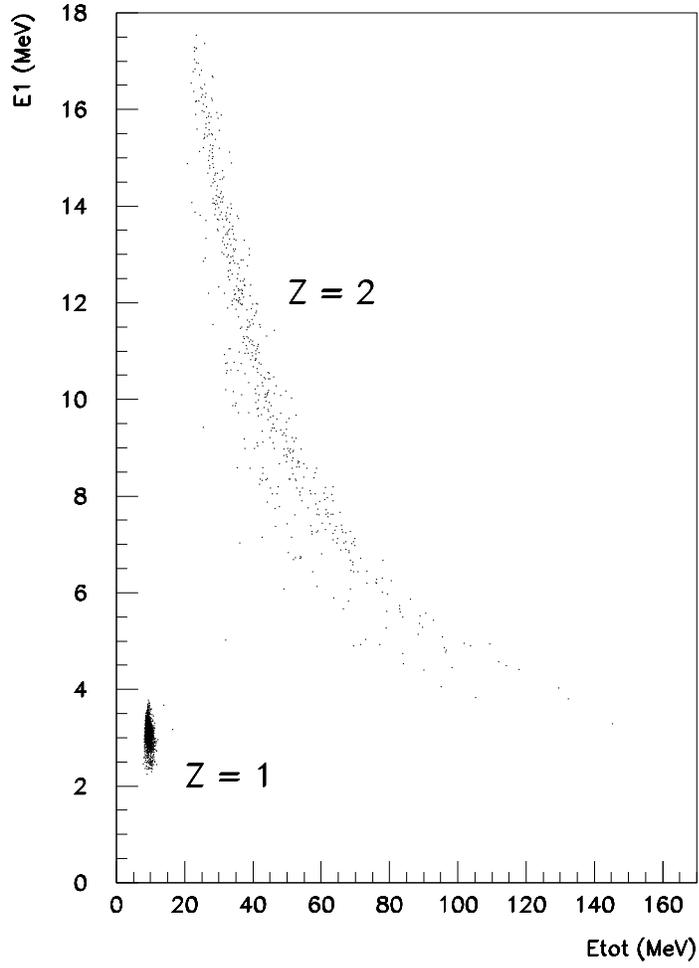}
   \caption{$E_1 \,\,\, vs \,\,\, E_{tot}$
 for events collected during the 1998 November 7 SEP event,
after the  action of the track filter. $E_1$ is the energy
released by  particles in the first silicon detector, and
$E_{tot}$ is the total energy released.}
\label{filtro}\end{figure}

The background reduction capability of the track selection
algorithm was previously tested with beam test data and in the
Galactic Cosmic Rays (GCR) analysis. Figure \ref{filtro} shows
$E_1 \,\,\,$ plotted against  $\,\,\, E_{tot}$, where $E_1$ is
the energy released by the particles in the first silicon
detector of the tower NINA, and $E_{tot}$ the total energy
released in the whole instrument, for events collected during the
1998 November 7 SEP event, after the application of the track
filter. It is evident that  almost all background was rejected,
whilst only eliminating  on average $\sim$ 3\% events from the
whole sample. This is evidence of small background contamination
in SEPs.

\vspace{0.7cm}

Charge and mass identification procedures are applied to events
which survive the track selection algorithm. The mass M and the
charge Z of the particles are calculated in parallel by two
methods, in order to have a more precise particle recognition: one
method uses the residual range, while the other uses  an approximation
of the Bethe-Bloch theoretical curve \citep{nina3}.

For a more complete  rejection of the background, only particles
with the same final identification according to the two methods
are selected.  Finally, a cross-check between the experimental
range of the particle in the detector and the expected value
according to the simulations gives a definitive consistency test
for the event. These consistency cuts eliminate about 0.1\% of
tracks detected during SEP events.

\subsection{Background estimations}

As anticipated above, in order to study  rare isotope abundances
($^{2}$H, $^{3}$H, $^{3}$He) in the analysis of SEPs it is
necessary to subtract the solar quiet background. Secondary
productions inside the instrument induced by high energy solar
particles must also be accounted for.

\vspace{0.5cm}

$\bullet$ {\it Solar quiet background}

This includes primary galactic  particles  (only $^{2}$H and
$^{3}$He), and secondary $^{2}$H, $^{3}$H and $^{3}$He which are
produced by nuclear interactions of primary cosmic rays inside
the $300 \; \mu m$ of aluminum which cover the first silicon
plane of NINA.

We measured this background component during passes over the
polar caps in solar quiet periods; the average counting rate was
$\sim (7.5\pm 0.9)\times 10^{-5}$ events s$^{-1}$ for deuterium,
$\sim (3.7\pm 0.6) \times 10^{-5}$ events s$^{-1}$ for tritium,
and $\sim (1.5\pm 0.1)\times 10^{-4}$ events s$^{-1}$ for
$^{3}$He, each of them relative to the energy interval reported
in Table \ref{thre}.

\vspace{0.5cm}

$\bullet$ {\it Secondary production by SEPs}

The amount of secondary productions inside the instrument,
induced by interactions of  solar particles,   cannot be directly
measured but can be inferred by estimations.

The most important reactions which can produce secondary $^{2}$H,
$^{3}$H   and $^{3}$He are the interactions of protons and
$\alpha$-particles with the aluminum cover:

\begin{enumerate}

\item  $ ^{27}Al \; (p,X) \; ^{2}H, \, ^{3}H, \, ^{3}He, \, ... $

\item  $ ^{27}Al \; (\alpha,X) \; ^{2}H, \,^{3}H, \, ^{3}He, \, ... $

\end{enumerate}

\noindent The ratio $R$ between secondary $^{3}$He  and primary
$^{4}$He nuclei, considering reaction 1. and 2.,  is given by:

\begin{equation}
\label{erre}
 R = \left( \frac{\sigma_{p} \:
F_{p}} {F_{\alpha}} \: + \: \sigma_{\alpha} \right) \:
\frac{N_{A}}{A} \: \rho \: \Delta x \,\, ,
\end{equation}

where $\sigma_{p}$ and  $\sigma_{\alpha}$   are the cross sections
for the reaction 1. and 2. respectively, and $F_{p}$ and
$F_{\alpha}$  the integral fluxes of incident protons and
$\alpha$ particles with sufficient energy  to produce a secondary
$^3$He detectable by NINA (E$_p > 40$ MeV, E$_{\alpha}
> 10$ MeV n$^{-1}$). $N_{A}$ is the Avogadro number,  $\Delta x$
the thickness of the aluminum cover, $\rho$  the density of
aluminum and $A$ its atomic weight.

To give  an estimation of the   $^{3}$He secondary production in
NINA, we adopt a  value  of  $\sigma_{p}$ = 30 mb
\citep{physrev,nim} and $\sigma_{\alpha}$ = 100  mb, as taken by
the authors in \citep{chen1995}. Eq. \ref{erre} then becomes:

\begin{equation} R = 2\times 10^{-4} \left(
0.3 \: \frac{F_{p}} {F_{\alpha}} \: + \: 1 \right) \,\, .
\label{bg}
\end{equation}

$F_{\alpha}$ is directly measured by our instrument. In order to
evaluate  $F_{p}$ we need to propagate the proton flux, that we
measure in the energy interval 12--14 MeV, to higher energies
utilizing the proton spectral index from
IMP-8\footnote{http://nssdc.gsfc.nasa.gov/space/space$_-$physics$_-$home.html.}.

Table \ref{phystable} reports  the values of  $R$  estimated for
the nine SEP events considered. The background coming from estimations by eq. 
\ref{bg}
is less than  10\%
 of the solar quiet background, except for one case (1998 November
 14) where an additional analysis was
 needed.

 \vspace{0.5cm}

\noindent Similar calculations of the ratio $R$ between secondary
$^{2}$H and primary $^{1}$H nuclei, and between secondary $^{3}$H
and primary $^{1}$H nuclei, give a result of $R$ $\sim$ $10^{-5}$.

\subsection{Dead time of the instrument}

During Solar Energetic events the counting rate  is orders of
magnitude higher than during solar quiet periods, so the dead time
of the instrument can have an important effect on  flux
measurements.

The time resolution of the trigger system in NINA is about 2
$\mu$s. If a particle produces a trigger, the signal from the
detectors is amplified and shaped before being sampled by an ADC.
The event is then stored in a FIFO for 2 ms. The data are read by
the on-board processor, which performs  pedestal subtraction and
zero suppression  for each channel,  and then  are stored in the
memory. This procedure takes 10 ms.

\begin{figure}
\epsscale{.7} \plotone{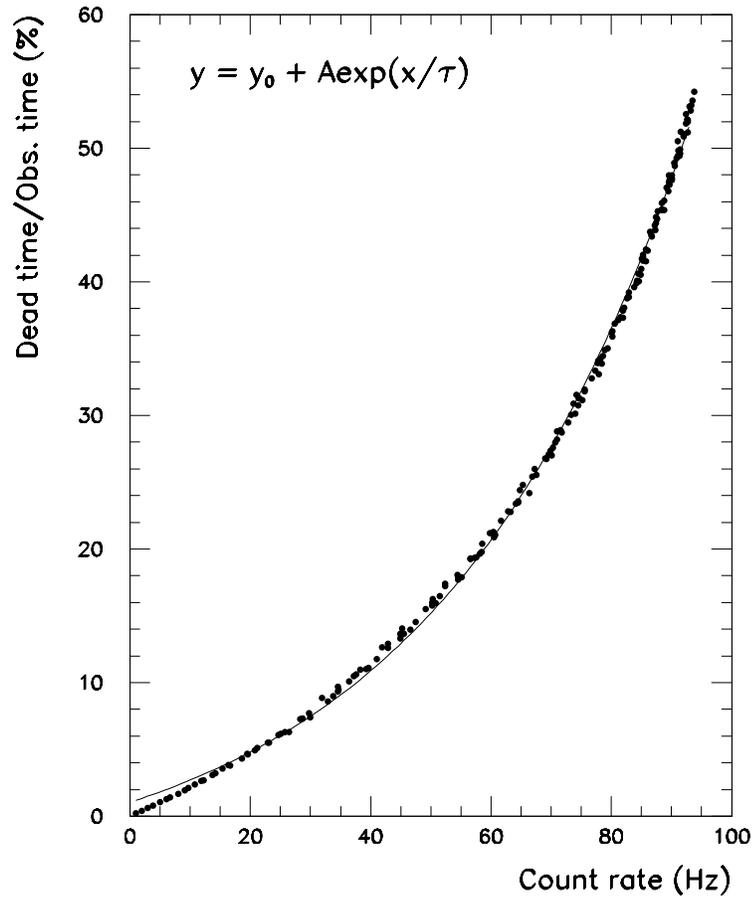}
   \caption{The ratio between the  dead time and the observation time of NINA, 
as a function
   of the external rate.
   An exponential fit has been super-imposed.}
\label{deadtime}\end{figure}

In order to estimate the dead time of the instrument, Monte Carlo
simulations  of the performance of the data acquisition system
were carried out, assuming a Poisson distribution of the particle
fluxes. Figure \ref{deadtime} shows the ratio between the
estimated dead time  and the observational time of the
instrument  as a function of the measured rate. This ratio is
fitted with an exponential function and  the parameters of the
fit have been used in the exposure time program (see \S 3.4). The uncertainty
on the fit parameters is of the order of 10\%, and this yields a
maximal inaccuracy of 2\% on the exposure time calculations for
all the SEP events analyzed.

\subsection{Flux measurements}

\begin{figure}
\epsscale{1} \plotone{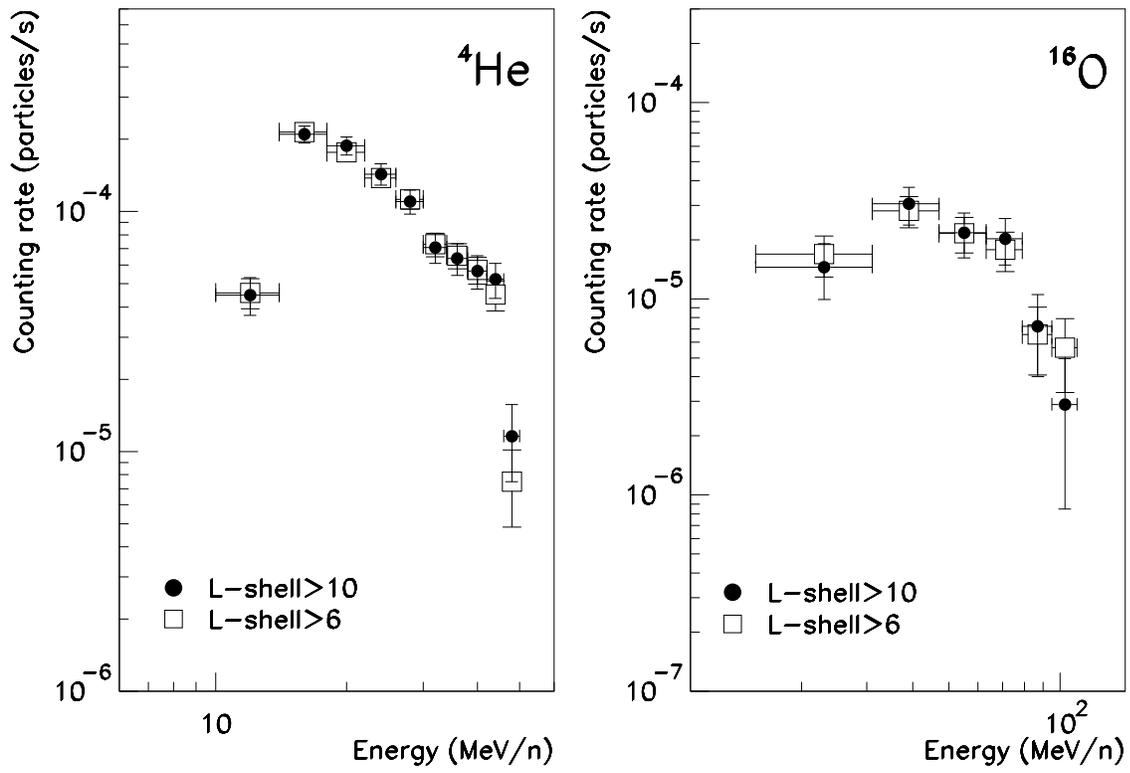}
   \caption{Counting rate, plotted as a function of energy,  for $^{4}$He (left) 
and $^{16}$O (right)
    measured at different geomagnetic latitudes.}
\label{lshell}\end{figure}

To be free from the effects of the rigidity cut-off due to the
Earth's magnetic field,  only events recorded in polar regions at
L-shell $> 6$ have been considered for SEP analysis. Figure
\ref{lshell} shows the counting rate of $^{4}$He and $^{16}$O, as
a function of energy, detected at two different geomagnetic
latitudes. The acquisition rate does not vary between the cut
L-shell $>$ 6 and $>$ 10.

\vspace{0.6cm}

Particle fluxes  were  reconstructed according to the following
formula:

\[ Flux(E)= \frac{\Delta N(E)}{ T \:  GF(E) \: \Delta E } \,\, , \]

where $\Delta N(E)$ is the number of  particles detected with an
energy between $E$ and $E+\Delta E$, $T$ is the  exposure time in
orbit, which takes into account also the dead time of the
instrument, $GF(E)$ is the average value of the Geometrical Factor
between $E$ and $E+\Delta E$ (see Figure \ref{gf}), and $\Delta
E$ is the relevant energy bin.

\section{SEP  measurements}

\begin{figure}
\epsscale{1.3} \plottwo{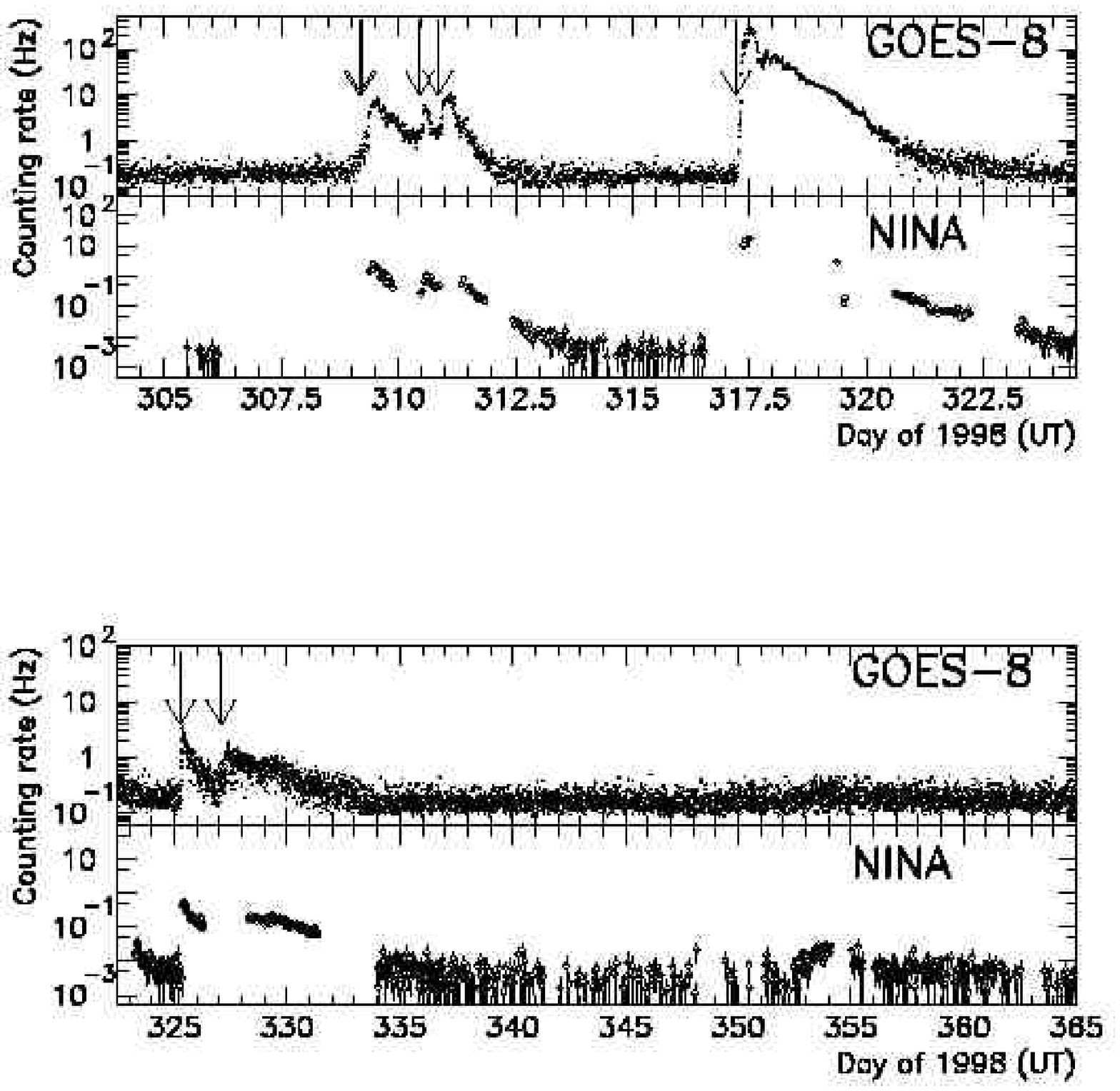}{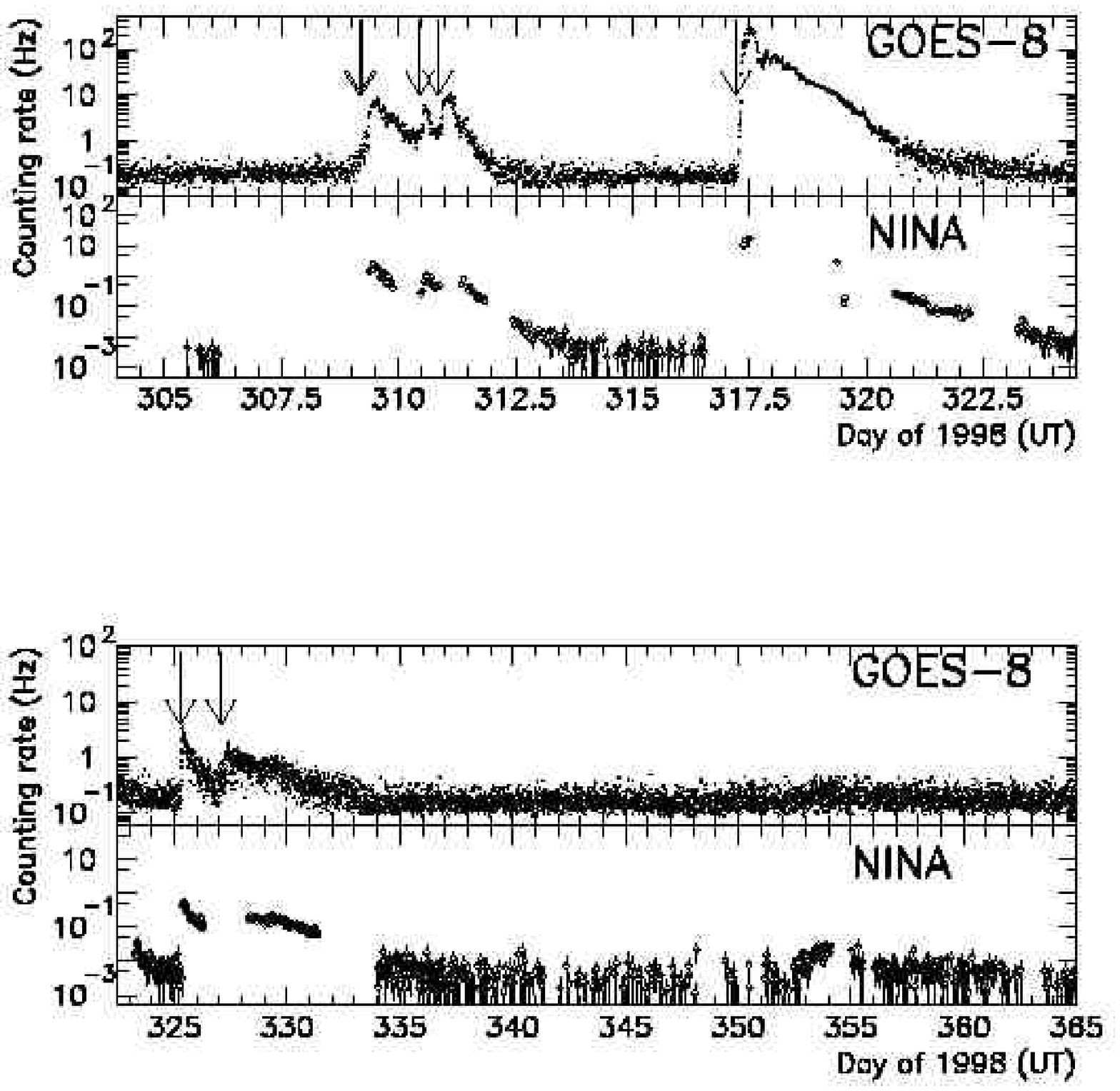}
   \caption{Proton  counting rate  for the period November 1998 -- March 1999, 
for GOES-8 (top)
   and NINA (bottom). Arrows mark the candidate X-ray  associated event, 
according to Table \ref{noa}.}
\label{9899prof}
\end{figure}

 SEP events
were identified by  an unpredictable increase in the trigger
 rate of the instrument,  at least one order of magnitude
with respect to the averaged solar quiet values. Nine such
increases in the period  October 1998 -- April 1999 have been
chosen for  analysis. To illustrate them, Figure \ref{9899prof}
shows the counting rate of protons with energy of $\sim$ 10 MeV
as a function of time measured by NINA in this period, together
with the GOES-8 proton intensity (energy $>$ 10
MeV)\footnote{http://www.sel.noaa.gov/Data/goes.html}.

A summary of  all nine events observed by NINA is presented in
Table \ref{noa}, together with characteristics of the solar
 events that can be associated to the
 SEPs\footnote{gopher://solar.sec.noaa.gov:70/11/indices.}.
 The event of 1998 November 14 was  the most powerful,
 where the counting rate increased of almost 3 orders of magnitude with respect
 to solar quiet periods, reaching a value of 70 Hz.  Figure \ref{9899prof}
indicates that the event lasted several days  and was detected by
our instrument in two separate emissions (Table \ref{noa}).
 For the other events
  we registered increases of one or two orders of magnitude on average.
The events of 1998 November 6-7-8 and those of 1999 January 20-22
occurred in a very close period of time, and there might be
effects of superposition between events. However, as it will be
shown later, their spectral characteristics and isotopic
composition are very different. For this reason we believe that
these SEPs may  have a different origin and we studied their
characteristics separately. As evident in Figure \ref{9899prof},
all chosen events are in good correlation with GOES-8
observations.

 \vspace{0.7cm}

\begin{figure}
\epsscale{0.9} \plotone{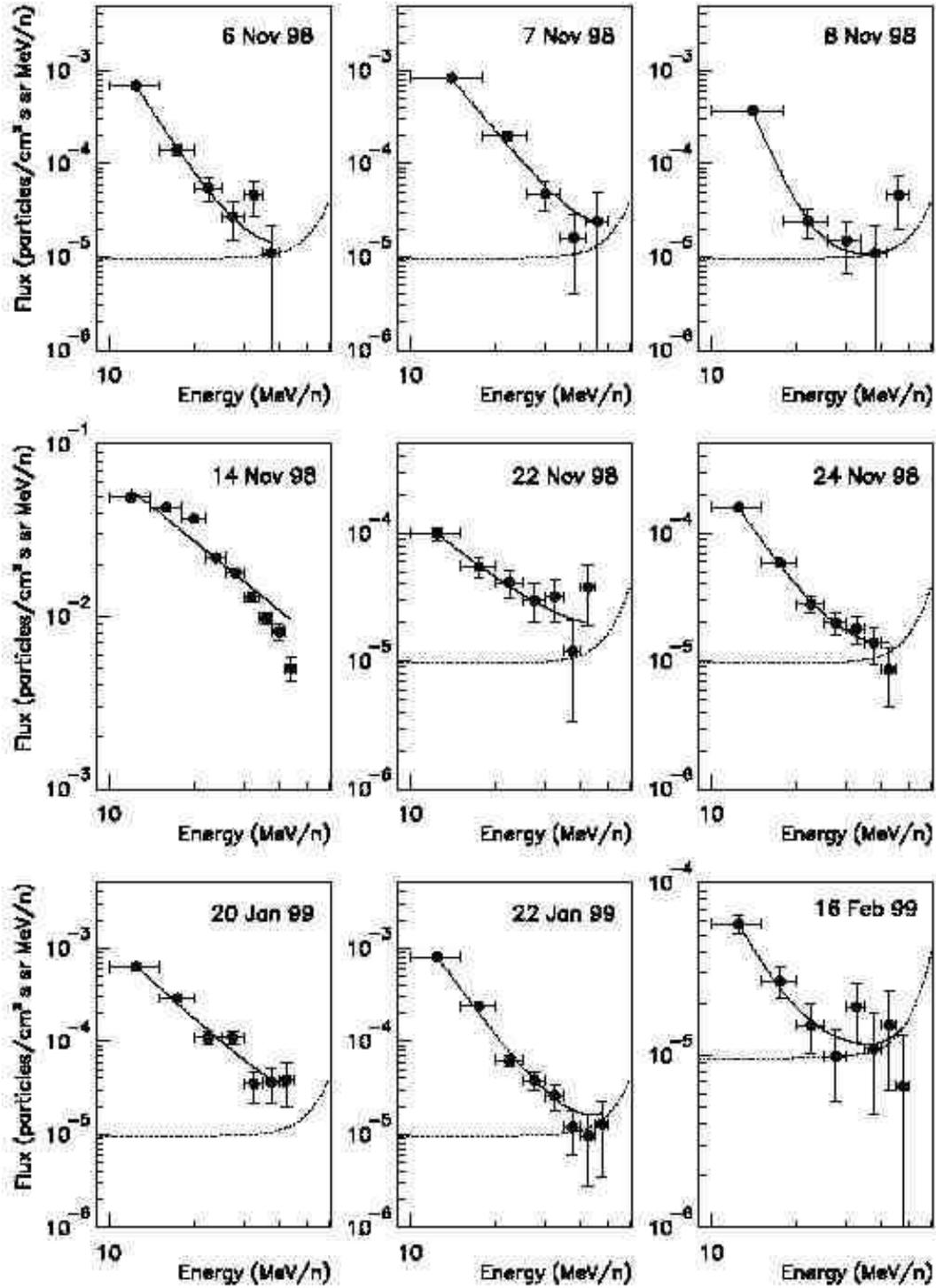}
   \caption{Differential energy spectrum for $^{4}$He for the 9  different SEP 
events
   detected by NINA. The dashed  line (see eq. \ref{be}) represents the 
background $B(E)$
   of galactic $^{4}$He;
   a power-law spectrum $S(E)$ (solid line, see eq. \ref{se}) has been super-
imposed to the data.}
\label{allhe4}\end{figure}

$^{4}$He observations range from  10   to 50 MeV n$^{-1}$. Figure
\ref{allhe4} presents the $^{4}$He fluxes that were reconstructed
during every SEP event. The dashed line on the spectra  is a fit
of the solar quiet galactic background of $^4$He, measured by our
own instrument \citep{nina3}. Data of the galactic background, in
our energy range, are well reproduced by the function:
\begin{equation}
B(E  [MeV \, n^{-1}]) = 9.6 \times 10^{-6} +  4.1 \times 10^{-9}
exp  (E/6.7) [cm^{2}\, sr\, s\,  MeV\, n^{-1}]^{-1}\,\,.
\label{be}
\end{equation}
The energy spectrum during the SEP
events   was  fitted
 by a power-law component plus the background (solid  line):
\begin{equation}
 S(E \: [MeV \, n^{-1}]) = A \: E^{-\gamma} + B(E) \;\;
[cm^{2} \: sr \: s \: MeV \: n^{-1}]^{-1} \,\,.
\label{se}
\end{equation}

The value of $\gamma$ (spectral index) for each event is reported
in Table \ref{phystable}. It varies considerably from event to
event,  ranging from 1.8 in the 1998 November 14 event to 6.8 in
the 1998 November 8 event. It is interesting to notice that the
1998 November 6 and 1998 November 7 events occur in the same NOAA
region (see Table \ref{noa}) but present different values of the
spectral index. The same holds for the 1998 November 22 and 1998
November 24 SEP events, in contradiction  to  observations by
\citet{chen1995} with the CRRES satellite, where events in the
same NOAA region tended to have similar spectral indexes.

\vspace{0.7cm}

Table \ref{phystable} summarizes  measurements of the ratio
($^{3}$He$_{SEP}$ - $^{3}$He$_{BG}$)/$^{4}$He in the range 15--45
MeV n$^{-1}$ for the nine SEP events, where $^{3}$He$_{SEP}$
represents the total flux of $^{3}$He which was measured during
the SEP events, and $^{3}$He$_{BG}$ the estimation of its overall
background, as discussed in section 3.2. During the 1998 November
7 and 14 SEP events this ratio is 3 standard deviations more than
the solar coronal value.

\vspace{0.7cm}

\begin{figure}
 \epsscale{0.8}
 \plotone{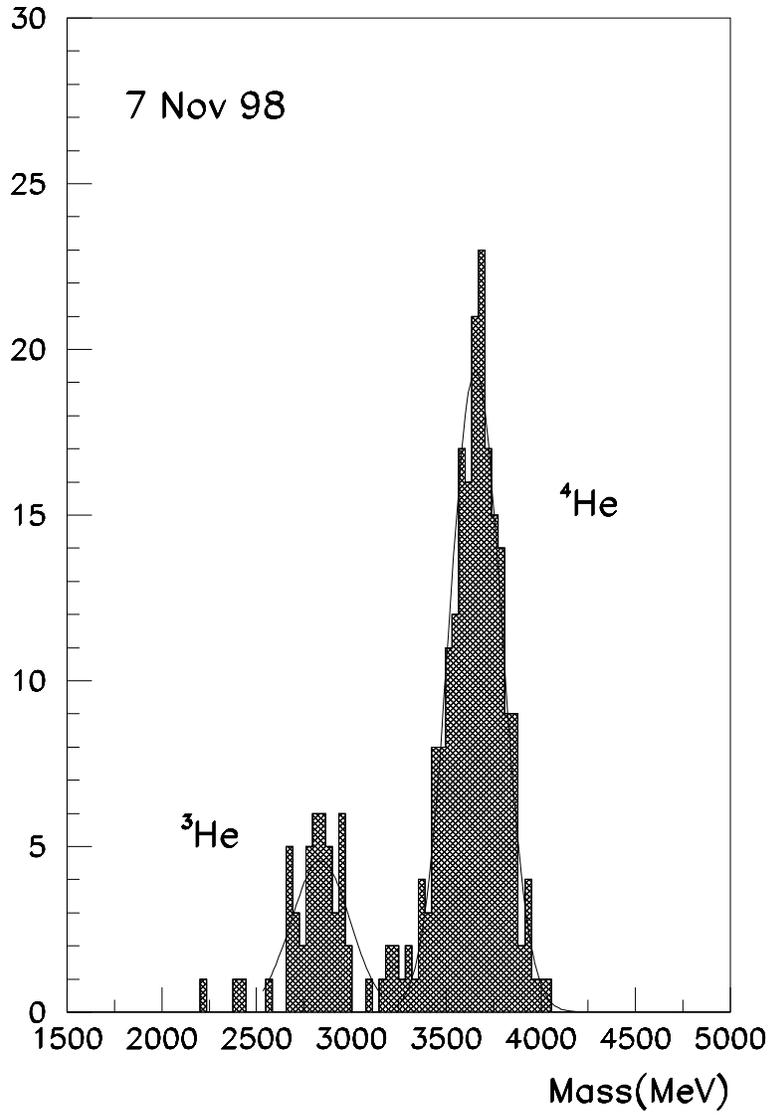}
\caption{1998 November  7 SEP event: mass reconstructions for
$^{3}$He and $^{4}$He.} \label{masse1}\end{figure}

\begin{figure}
\epsscale{0.8}
 \plotone{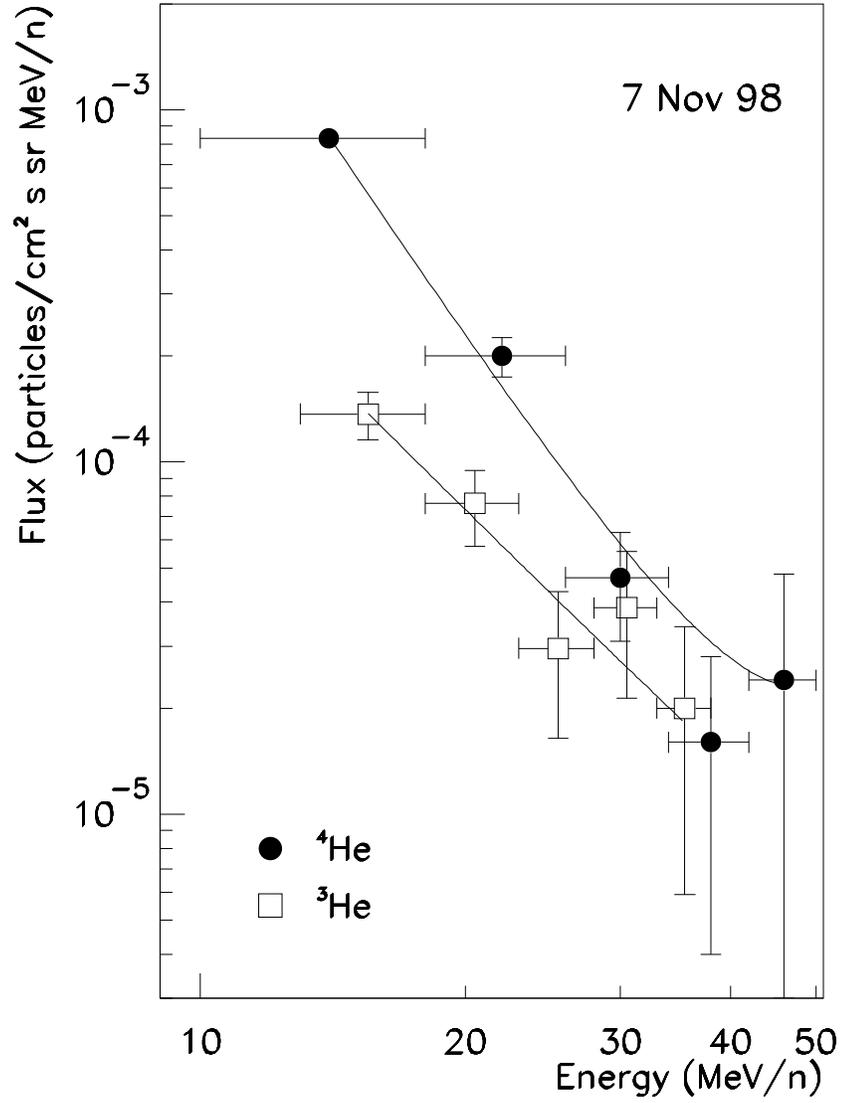}
\caption{1998 November 7 SEP event: differential energy spectrum
for $^{3}$He and $^{4}$He. } \label{he3he41}\end{figure}

For the 1998  November  7 SEP event we reconstructed the masses
of $^{3}$He and $^{4}$He (Figure \ref{masse1}), and  it was
possible to plot the $^{3}$He  differential energy spectrum over a
wide energy interval  (Figure \ref{he3he41}).  The $^{3}$He
spectrum, also fitted by a power-law,  is slightly harder
($\gamma$ = 2.5 $\pm$ 0.6) than that of $^{4}$He ($\gamma$ = 3.7
$\pm$ 0.3). This implies that the $^{3}$He/$^{4}$He ratio
increases with energy in this event. This tendency is also
confirmed by a comparison between our data and measurements taken
on board ACE [ULEIS instrument, \citep{mason1999}] over the energy
range 0.5--2.0 MeV n$^{-1}$ and averaged over the period 1998
November 6--8. Extrapolating  the $^{3}$He energy spectrum
measured by NINA to lower energies, the inferred $^{3}$He/$^{4}$He
 ratio for this SEP event would be about $10^{-4}$, in agreement with
 their value and much lower that what we obtain at higher energies.

\begin{figure}
\epsscale{1.1}
 \plotone{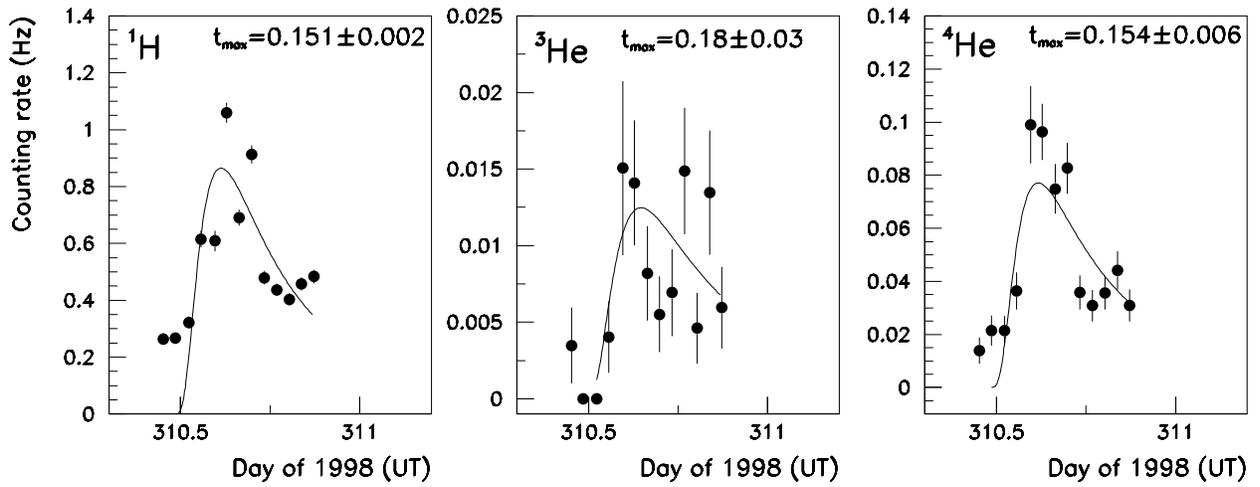}
\caption{1998 November 7 SEP event: time profile of the counting
rate of $^{1}$H (left),  $^{4}$He (centre) and $^{3}$He (right)
during the SEP event. t$_{max}$ (days) is the fit parameter
appearing in eq. \ref{prof}, which corresponds to the maximum  of
the time profile. } \label{timeprofile}
\end{figure}

For the 1998 November 7  SEP event we  reconstructed the emission
time profiles of different nuclei. Figure \ref{timeprofile} shows
the time profile of the counting rate of $^{1}$H, $^{3}$He and
$^{4}$He detected over the polar caps during this event. Every
point on the plots corresponds to a passage over the pole, which
lasted  about 10 minutes. The profiles have been fitted by the
following function, which takes into account both  propagation in
the solar corona and the diffusion of particles to the Earth
\citep{burlaga}:

\begin{equation} Counting \,\, \, rate\,\,\, [t (days)] = A \,(t-t_0)^{-2.5}
\,\,\, e^{-2.5\,t_{max}/(t-t_0)} \,\,\, Hz,
\label{prof}
\end{equation}

where t is expressed in days (UT), $t_0$ corresponds to the
beginning of the 1998  November 7 event, and A and t$_{max}$ are
the two parameters of the fit. Direct propagation from Sun to
Earth takes, for 10 MeV n$^{-1}$ particles, roughly 1 hour. Taking
into account this value, and assuming that the main part of the
particle emission occurs before t$_{max}$, we estimate the time
for coronal propagation  for the three nuclear species to be not
more than  3 hours. This similarity  suggests the same
acceleration and transport mechanisms for the nuclear species,
despite their different charge-to-mass ratio.

\vspace{0.7cm}

\begin{figure}
 \epsscale{0.8}
 \plotone{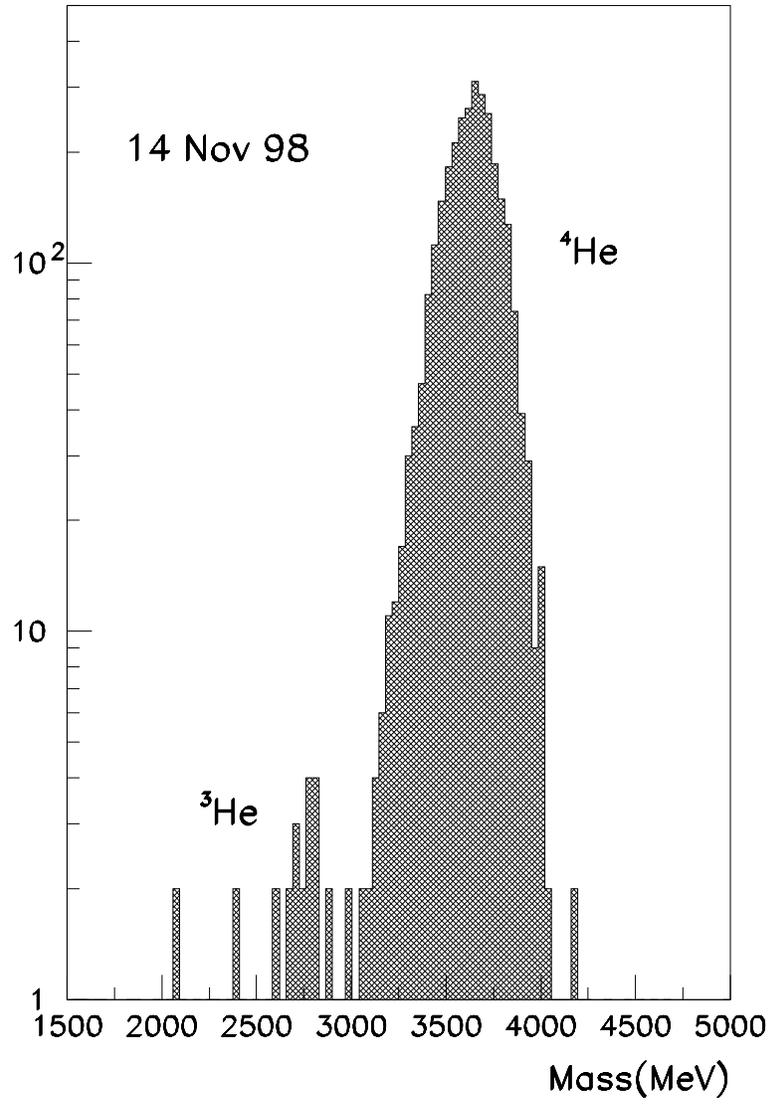}
\caption{1998 November 14 SEP event: mass reconstructions for
$^{3}$He  and $^{4}$He (Energy $>$ 25 MeV n$^{-1}$).}
\label{masse2}\end{figure}

\begin{figure}
 \epsscale{0.8}
 \plotone{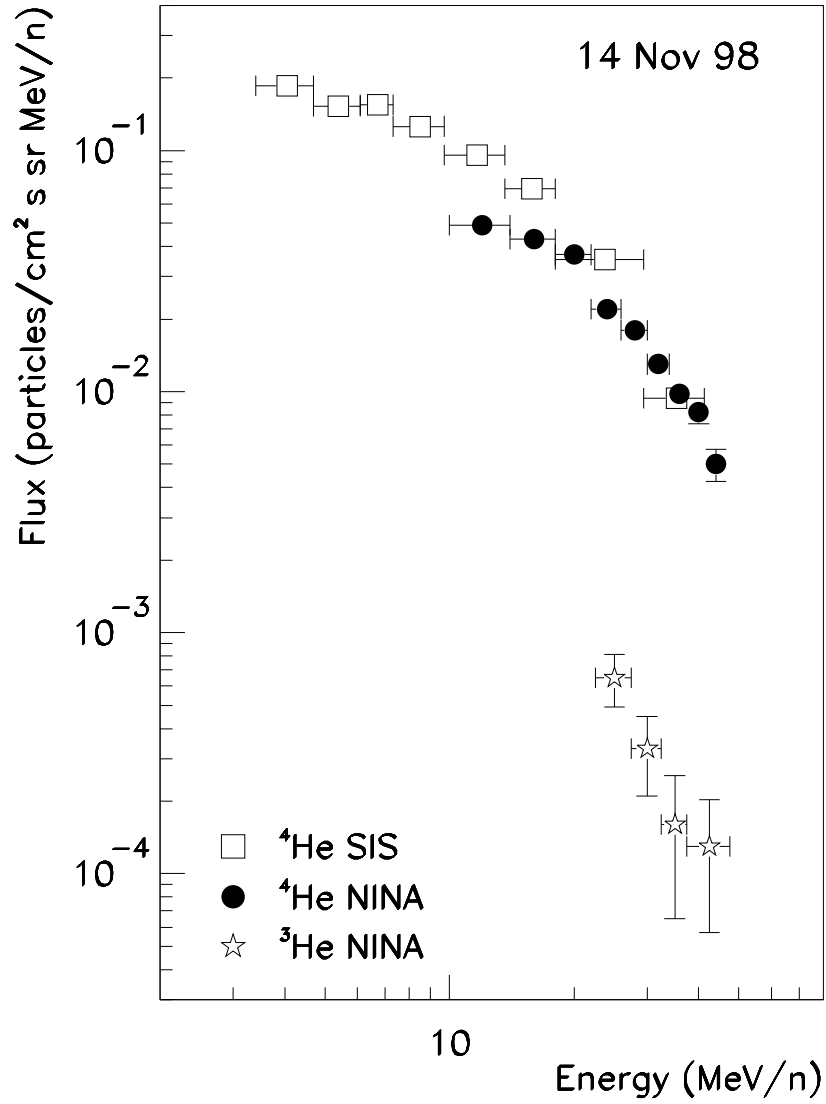}
\caption{1998 November 14 SEP event:  differential energy
spectrum for $^{3}$He (Energy $>$ 25 MeV n$^{-1}$) and $^{4}$He.
The white squares are data from SIS (ACE satellite).}
\label{he3he42}\end{figure}

The strongest solar event that we detected, as already mentioned,
was the one of 1998 November 14. Due to the very high flux
intensities, which increased the noise of the detector,  the
$^{3}$He spectrum was reconstructed only by nuclei which crossed
at least 7 silicon layers in the instrument, so having energies
greater than 25 MeV n$^{-1}$. Figure \ref{masse2} shows the
helium isotope mass reconstructions for tracks with at least 7
views hit.  Figure \ref{he3he42} presents the energy spectrum of
$^{3}$He and $^{4}$He together. On the same figure  $^{4}$He
measurements from SIS on board ACE are also
reported\footnote{http://www.srl.caltech.edu/ACE/ASC/level2.}.
For the 1998 November  14  event there are other measurements of
the $^{3}$He/$^{4}$He ratio, performed by the IMP-8
\citep{dietrich1999} and SIS instrument \citep{cohen1999}. Their
measured values are  R = 0.02 $\pm$ 0.01 in the energy interval
30--95 MeV n$^{-1}$, and R = 0.005  in the range 8--14 MeV
n$^{-1}$.

\vspace{0.5cm}

\begin{figure}
\epsscale{1.}
 \plotone{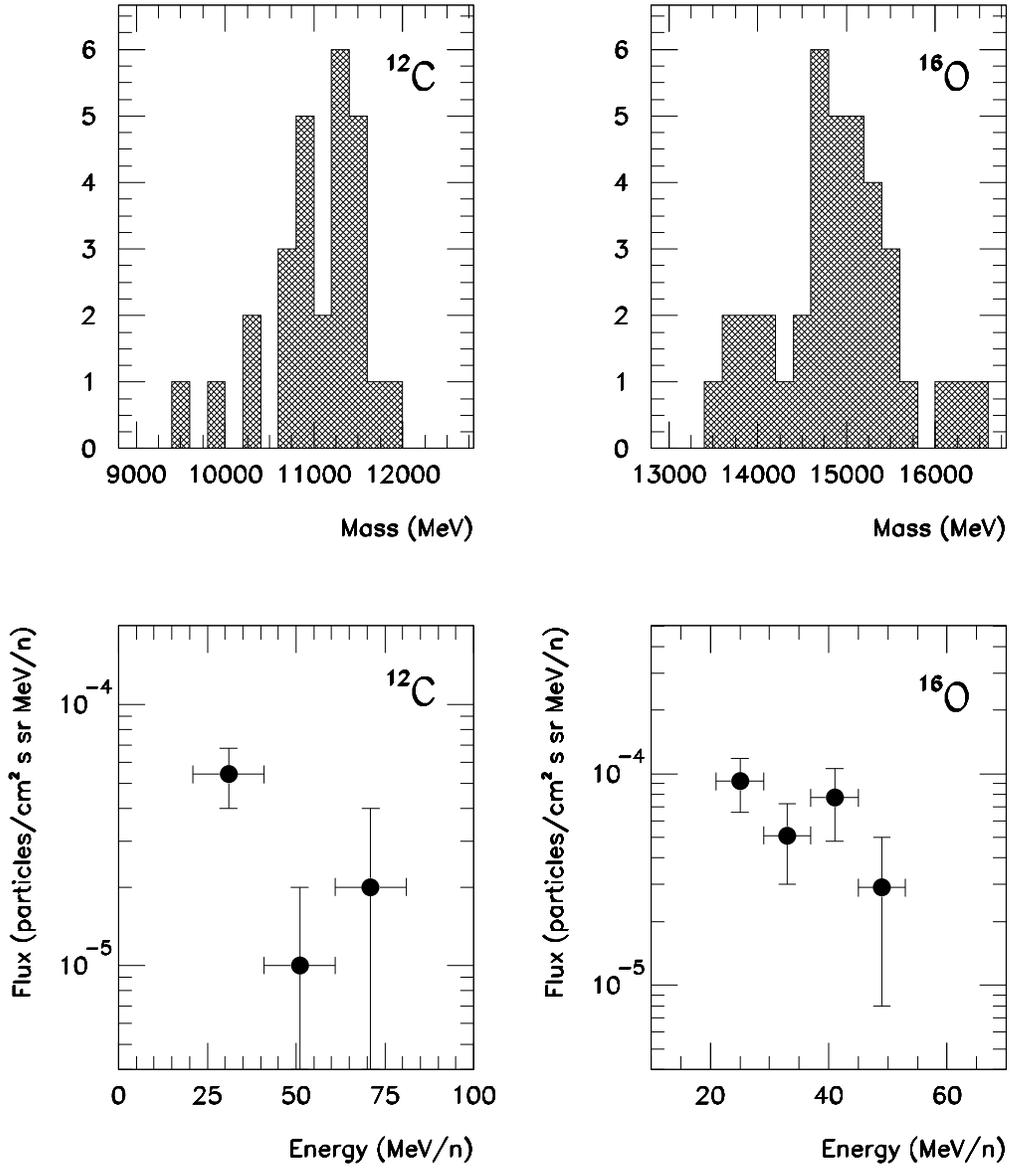}
\caption{1998 November 14 SEP event. Top: mass reconstruction for
$^{12}$C and $^{16}$O. Bottom: differential energy spectra for
$^{12}$C and $^{16}$O.} \label{co}\end{figure}

 During the 1998  November  14 SEP event  there was also a strong presence of
heavy elements. Figure \ref{co} presents the  mass reconstructions
and the differential energy spectra for carbon and oxygen.

\vspace{0.8cm}

The High Threshold mode allowed the observation of hydrogen with
its isotopes  over the narrow energy windows shown in Table
\ref{thre}. Table \ref{phystable} presents the ratios between the
deuterium and proton fluxes  after background subtraction, in the
range 9--12 MeV n$^{-1}$.   Since the two isotope measurements
span two different energy regions (see Table 1), we measured the
proton flux  in the range 11-16 MeV and utilized the proton
spectral index from
IMP-8\footnote{http://nssdc.gsfc.nasa.gov/space/space$_-$physics$_-$home.html.}
to extrapolate the proton flux to the deuteron energy region. The
ratio $^{2}$H/$^{1}$H has an average value of about $(3.9 \pm
1.4)\times 10^{-5}$ for all events; this value is in agreement
with a previous measurement \citep{anglin1975}, which reported a
$^2$H/$^1$H ratio  equal to $(5.4 \pm 2.4)\times 10^{-5}$ between
10.5 and 13.5 MeV n$^{-1}$, when averaged over  a large number of
SEP events, consistent with solar abundance values
\citep{1986mcguire}.

\begin{figure}
\epsscale{1.} \plotone{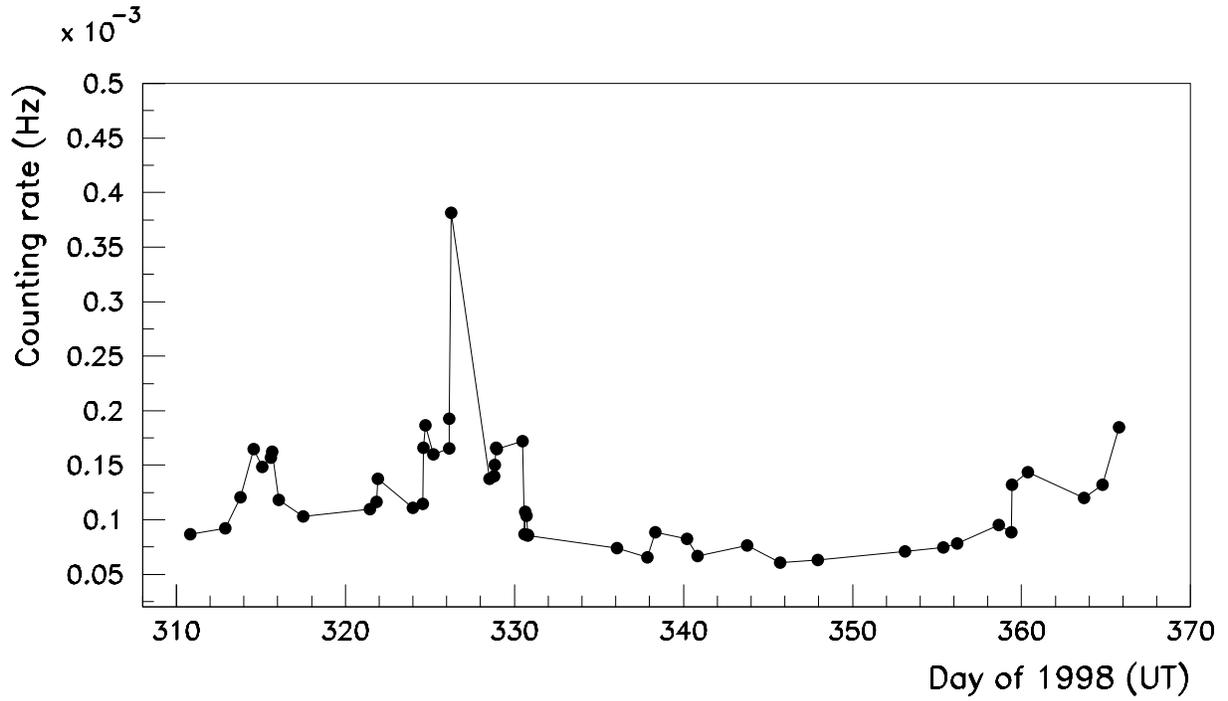} \caption{Counting rate of
deuterium (Energy $>$ 9 MeV n$^{-1}$) during the months
November--December 1998.} \label{deuprofile}\end{figure}

 In the 1998
November 24 event  the deuterium emission was more intense, with
a deuterium-to-proton ratio   equal to $(3.5 \pm 1.4)\times
10^{-4}$, almost 10 times higher than the coronal value. Figure
\ref{deuprofile} presents the counting rate of deuterium during
the months November--December 1998. To produce this figure,
 a technique that first calculates the time interval
$\Delta$T covering a fixed number N of successive events of
deuterium was used.  The counting rate is then equal to:
(N-1)/($\Delta$T-$\tau$), where $\tau$ is the dead time which also
includes the time spent outside the polar regions. The counting
rate  is determined sequentially for each event, and assigned to
the time when  the first event was recorded \citep{efre1997}.  In
Figure \ref{deuprofile} we have chosen N=7. Due to background
conditions two days, in correspondence  to the 1998
 November  14 event, were excluded from the analysis.

\begin{figure}
 \epsscale{1.}
 \plotone{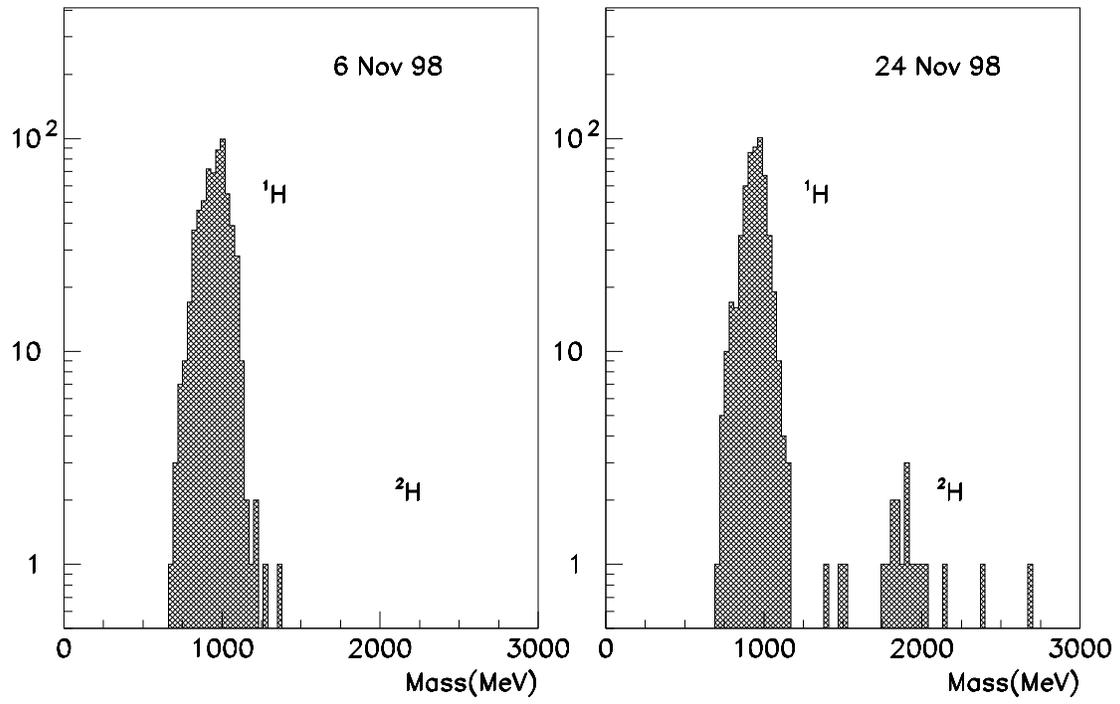}
\caption{1998  November 6 and 1998  November  24 SEP event: mass
reconstruction (Energy $>$ 9 MeV n$^{-1}$) for the hydrogen
isotopes.} \label{deuterio}\end{figure}

In Figure \ref{deuprofile} a peak of counting rate is visible, in
possible correlation with the   1998  November 24 SEP event.
Figure \ref{deuterio} shows the mass reconstruction of hydrogen
isotopes for this event (right) together with the distribution of
the 1998 November 6 event (left). These two SEP events correspond
 to the maximum and the minimum of the deuterium
detection respectively.

\vspace{0.5cm}  From our data it was possible to estimate upper
limits for tritium. The last column of Table \ref{phystable}
presents the ratio of counting rates of $^{3}$H in the interval
6--10 MeV n$^{-1}$, and of $^{1}$H in the range 10--14 MeV.

\section{Discussion}

The isotope abundance ratio  and the energy spectra of SEPs are
related to the acceleration mechanisms and propagation processes
involved.

Among the nine $^3$He/$^4$He ratios  calculated from our SEP
events, only the 1998 November 7  one is clearly above the limit
$^3$He/$^4$He $>$ 0.1, which is the  the boundary value separating
gradual from impulsive events \citep{Reames1999}. The energy
spectra and counting rate  profiles of $^3$He and $^4$He during
this event were presented  in Figure \ref{he3he41} and
\ref{timeprofile}, and partially discussed in the previous
section. By using the IMP-8 spacecraft
data\footnote{http://nssdc.gsfc.nasa.gov/space/space$_-$physics$_-$home.html.}
it is possible to also extract  the average value of the proton
spectral index, $\gamma$,
 equal to  (3.6 $\pm$ 0.1) in the energy range
11--74 MeV. This value is practically the same as the $^4$He
spectral index (see Table \ref{phystable}) measured by NINA for
this SEP event.

The energy spectra of Solar Energetic Particles measured at 1 AU
can be modified, with respect to the emission spectra, by
propagation effects. The absolute value of the mean free paths of
particles traveling from the Sun to the Earth can vary by one
order of magnitude between individual events, but, as shown in
\citet{droge2000}, the shape of the rigidity dependence does not
vary. This rigidity dependence  is a power-law with a slope equal
to 0.3 in NINA's rigidity intervals. Since the $^{3}$He rigidity,
at the same energy per nucleon, is between that of proton and
$^{4}$He, it would be difficult in the 1998 November 7 SEP event
to explain the hardness of the $^{3}$He spectrum compared to that
of $^4$He and  of protons only with diffusion effects from the
Sun. The differences in the spectral shapes of $^{3}$He and
$^{4}$He, observed at 1 AU in this event, most probably   already
existed during the SEPs emission.

Models of stochastic acceleration of Solar Energetic Particles
near the Sun predict a monotonically increase in the $^3$He/$^4$He
ratio with the energy, as we observe in this SEP event,  with
rigidity-dependent spatial and energy diffusion coefficients
\citep{moe1982,mazur1992}. Any model dependence on the particle
rigidity  effects  the spectral shapes of particles according to
their charge-to-mass ratio. This is not  observed in the 1998
November 7 event. Propagation effects in the corona appear to be
small in this case and the differences between the energy spectra
of $^3$He and $^4$He could reflect differences in
 the spectra at the source.

Acceleration processes based on the shock drift mechanism can
accelerate particles up to 100 MeV n$^{-1}$. Such models predict a
power-law energy spectra of ions with a mean value of the
spectral index between 3.5 and 4.5  \citep{anas1994}. In the
framework of this model it would be hard to explain the
differences  observed  between the $^3$He and $^4$He energy
spectra and especially the low value of the  $^3$He $\gamma$
index in this SEP event.

Wave resonance is hence the most likely mechanism for $^{3}$He
acceleration in the 1998 November 7  SEP event \citep{roth1997}.
This mechanism accelerates both $^3$He and heavy ions. The
abundance of heavy and $^3$He ions is determined by the
temperature and density of the flare plasma, and by the wave
properties. It is interesting to note that data reported by ACE
\citep{klecker1999} identify this SEP event as  gradual by the
low  Fe/O ratio. In this work  the ionic charge of several heavy
ions, including iron, was determined. These measurements at low
energy (0.2--0.7 MeV/n) are consistent with an equilibrium plasma
temperature of $\sim$ 1.3--1.6 $\times 10^6$ K and with typical
solar wind values, suggesting acceleration from a solar wind
source. With a plasma temperature of $\sim$ 2 MK \citet{roth1997}
predict the existence of a large population of oxygen at the same
energy per nucleon as $^3$He, which was not observed by our
instrument. Another possible reason for the discrepancy between
this event characteristic measured by the two instruments is the
narrow cone of emission for particles from $^3$He-rich events
\citep{reames1991}.
 It would be interesting to confront our
measured spectral behaviour of $^3$He and $^4$He with the
predictions of this model, but complete spectral calculations are
not yet available.

\vspace{0.9cm}

Some observations reported, already since  1970
\citep{hsieh,webber1975}, that a small $^3$He enrichment is also
present in large  events. Due to instrumental limitations of most
of the experiments in space, however, systematic $^3$He/$^4$He
measurements in large events with small $^3$He enrichment were
not available. More recently, \citet{chen1995} analyzed 16 SEP
events and found that even extremely large SEP events had a value
of $^3$He/$^4$He greater than 0.5\% - one order of magnitude
greater than  solar wind values. This evidence was reported also
by \citet{mason1999}, who observed 12 large SEP events with an
average value of $^3$He/$^4$He about ${(1.9\pm 0.2) \times
10^{-3}}$.

To explain this enrichment, \citet{mason1999} suggest that
residual $^3$He  from impulsive SEP events forms a source material
for the $^3$He  seen in some large SEP events.  In such cases, the
$^3$He  is accelerated out of this interplanetary suprathermal
population by the same process that energizes other particles at
these energies, namely the interplanetary shock.

Our $^3$He/$^4$He ratio measurements, reported in Table
\ref{phystable}, seem to confirm the fact that in all SEP events
there is a quantity of $^3$He greater than that typical of solar
coronal values ($^3$He/$^4$He about ${4\times 10^{-4}}$). If we
average the $^3$He/$^4$He ratio over all events detected by NINA,
except the 1998 November 7--8 which is  clearly $^3$He-enriched,
we obtain the value $(1.1 \pm 0.4)\times 10^{-2}$.

Nuclear interactions in acceleration regions  can also contribute
to the high value of the ratio  $^3$He/$^4$He, with respect to
coronal values. As stated in the introduction,  at flare physical
sites the possible presence of secondary particles from nuclear
interactions \citep{ramaty1987} should not be excluded. Such
interactions also produce  $^3$H and $^2$H, which are the
signature of the process. From $\gamma$-ray spectroscopy there is
evidence for the presence of $^3$He and $^2$H at the flare sites
\citep{mand1999,chupp1983,tere1993}. Usually all secondary
particles are trapped in the flare loops but some of them can
escape into the interplanetary space.

Particles escaping from Sun during acceleration traverse a small
 amount of material and do not undergo many interactions. It is
therefore difficult to separate solar deuterium particles from
the galactic and instrumental background. In our measurements the
values of the deuterium flux are close to the instrumental limit.
For most of the SEP events there was not an appreciable quantity
of $^2$H, and in some cases we could determine only upper limits
on the ratio $^2$H/$^1$H  (see Table \ref{phystable}). The average
value of this ratio, over all our events, is equal to $^2$H/$^1$H
$ = (3.9\pm 1.4)\times 10^{-5}$, which corresponds to about 0.1 g
cm$^{-2}$ thickness of traversed material, for protons with
energy about 30 MeV  \citep[Figure 10 therein]{ramaty1974}. This
thickness gives a $^3$He/$^4$He ratio of about 10$^{-3}$, that is
lower than our average value. However, it should be noted that
the $^2$H/$^3$He ratio measured by probes in the interplanetary
space may be depressed by a factor 10 due to differences in $^2$H
and $^3$He angular distribution and propagation processes
\citep{colgate1977}. A theoretical analysis of this ratio depends
on the solar flare model used,  which is still very
controversial, and  beyond the scope of this paper.

In 1998 November 24 SEP  event the $^2$H/$^1$H ratio  was about
10 times higher than the solar corona abundance ratio, and also
above the secondary production estimations. This quantity of
deuterium cannot be explained by the {\it thin target model}
because the amount of estimated traversed material would stop a
5--10 MeV n$^{-1}$ deuterium due to energy losses. Despite the
secondary production estimations suffer from uncertainties, due
to the uncertainty in the  cross sections and possible
fluctuations in the galactic background, we have some arguments
that exclude an under-estimation of the background and confirm a
solar origin for the measured deuterium:

\begin{itemize}

\item[(i)] the peak in $^2$H count rate in the 1998 November 24 SEP  event
does not correspond to a peak in the high energy proton and
$\alpha$ count rates. Such particles are the main source of
background production in our instrument;

\item[(ii)] the $^3$He/$^4$He ratio  measured by NINA in the 1998
November 6, November 14 and 1999 January 20 SEP events are
consistent with other measurements \citep{mason1999,cohen1999}.
This suggests that we do not have a higher level of secondary
$^{3}$He background, and therefore of deuterium, in our
measurements.

\end{itemize}

  Following  the hypothesis reported by \citet{mason1999}
  it could be
 suggested that the deuterium observed by NINA on   1998 November 24
was produced in previous impulsive events, and  remained in the
low corona before being  erupted and accelerated,  in the same
way as
 $^3$He. This suggestion is supported  by observations of the Tibet solar 
neutron
 telescope, which
 observed possible  solar neutrons in association
with the  flare of 1998 November 23 \citep{hoshida1999}. It is
known that the mechanism  producing neutrons also concerns
deuterium synthesis. In association with this, \citet{yoshi1999}
reported gamma ray lines observed during the 1998 November 22
event at the same NOAA region as for our 1998 November 24 SEP
event. Taking into account these facts we suggest that some
deuterium was produced during November 22 and 23, a part of it
erupted, was accelerated and arrived at 1 AU in the next days.

In secondary production, the $^2$H/$^3$He ratio is expected to be
between 1 and 2, as inferred from data in the work by
\citet{colgate1977}. The angular dependence of secondary products
is not important because the time between production and emission
is long enough to make the flux of secondaries isotropic.
 In the 1998
November 24 event a $^3$He enrichment was perhaps also present
(Table \ref{phystable}). The ratio $^2$H/$^3$He measured by NINA
was of order unity, as expected. Unfortunately, due to limited
statistics, we cannot study this ratio in more detail.

In conclusion, the presence of deuterium in SEPs, coming from
secondary interactions in the solar ambient, suggests that part
of the $^3$He contents in large  Solar Energetic Particle events
may also have this origin.

\section{Summary}

We have determined the $^3$He/$^4$He ratio and helium energy
spectra over the energy range 10--50 MeV n$^{-1}$ for 9 SEP events
measured in the period October 1998 -- April 1999 by the
instrument NINA. The most interesting of these events was
recorded on 1998 November 7, where the ratio reached a value of
about 30\% but also presented  features typical of a "pure"
gradual event \citep{klecker1999}. The similarity of the time
profiles for the $^1$H, $^3$He and $^4$He emissions in the event
implies that these isotopes underwent the same acceleration
mechanism.

The other SEP events yield an average value of the $^3$He/$^4$He
ratio slightly higher that that typical of the solar wind. During
the 1998 November 24 events we  most probably detected solar
deuterium, with a $^2$H/$^1$H ratio   about 10 times more than
expected solar values in the energy range 9--12 MeV n$^{-1}$. The
average value of the $^2$H/$^1$H ratio, over all our events, is
equal to $^2$H/$^1$H $ = (3.9\pm 1.4)\times 10^{-5}$. The
presence of deuterium in SEP events, coming from secondary
interactions in the solar ambient, suggests that part of the
$^3$He contents in Solar Energetic Particles may also have this
origin.

\acknowledgments{V. Mikhailov wishes  to thank personally Prof. O.
Terekhov and Prof. D. Samarchenko for the many hours of fruitful
discussion. We acknowledge  the Russian Foundation of Base
Research, grant 99-02-16274, who partially supported the Russian
Institutions for this work.}

\clearpage

\begin{deluxetable}{ccc}
\tablecolumns{3} \tablecaption{Energy windows  for contained
particles in  High Threshold Mode \label{thre}} \tablewidth{0pt}
\tablehead{ \colhead{Particle}&\colhead{Z}&\colhead{Energy
window} \\ & & (MeV
 n$^{-1}$)} \startdata
 $^1$H  &  1&   11 - 16  \\
 $^2$H  &  1&   7 - 13\\
 $^{3}$H   &1 & 5 - 12\\
  $^3$He  &  2&   12 - 58 \\
 $^4$He& 2   &10  -   50   \\
 $^{12}$C&6 &18 - 90  \\
 $^{16}$O &8&   21 -107\\
 \enddata
\end{deluxetable}

\begin{deluxetable}{ccclcc}
\scriptsize \rotate \tablewidth{15cm} \tablecolumns{5}
\tablecaption{SEP events and characteristics of associated solar
events \label{noa}} \tablecomments{Second column: NINA
observation time (day of the year) for the SEP event.  Third
column: NOAA region number of the associated flare. Fourth
column: importance of the flare in terms of X-ray/H$\alpha$
classification. Fifth column: location of the flare in
heliocentric coordinates.  Sixth column: starting time (hh:mm) of
the X-ray event.}
\tablerefs{gopher://solar.sec.noaa.gov:70/11/indices.}
\tablehead{ \colhead{SEP date}&\colhead{Observation
time}&\colhead{NOAA }& \colhead{Class}&\colhead{Location}&
\colhead{Time of X-ray}
 \\ &  & & (X-ray/H$\alpha$) &  & {
 event (UT)}} \startdata

6 Nov. 1998& 309.38 $\div$ 310.52  & 8375&C1.1 / SF &N19W25& 04:38\\
   &&& C1.4 / SF && 04:56\\
7 Nov. 1998 & 310.52 $\div$ 310.89 & 8375&M2.4 /-&-& 11:06\\
8 Nov. 1998 & 311.34 $\div$ 311.87  & 8379 &C2.4 / SF
&S20W67&20:20
   (7 Nov.)\\
14 Nov. 1998&  317.36 $\div$ 317.54 & 8385&C1.7 / BSL&N28W90&
  05:18 \\
   &   320.60 $\div$  322.25 & && \\
22 Nov. 1998 & 325.32 $\div$  326.33 & 8384&X3.4 / 1N&S27W82&
   06:42 \\
24 Nov. 1998 & 328.30 $\div$  331.14   & 8384&X1.0 / -&-& 02:20 \\
20 Jan. 1999&  19.96 $\div$  21.93 & - &M5.2 / - & - & 20:04 \\
22 Jan. 1999& 21.93 $\div$  23.90  & 8440&M1.4 / SF &N19W44&
   17:24 \\
16 Feb. 1999& 46.23 $\div$  48.04 & 8458&M3.2 / SF &S23W14& 03:12\\

\enddata
\end{deluxetable}
\clearpage

\begin{deluxetable}{ccccccc}
\rotate
\tablewidth{0pt}
 \tablecolumns{7}
\tablecaption{Summary of physics features of SEP events
\label{phystable}} \tablehead{ \colhead{SEP date}&
\colhead{$\gamma$ \tablenotemark{\S} } & \colhead{$^{4}$He/$^{1}$H
\tablenotemark{\flat} } & \colhead{R($^{3}$He/$^{4}$He) } &
\colhead{$^{3}$He/$^{4}$He \tablenotemark{\dag}} &
\colhead{$^{2}$H/$^{1}$H \tablenotemark{\sharp}} &
\colhead{$^{3}$H/$^{1}$H
\tablenotemark{\ddag}} \\
& & & & $\times 10^{-2}$ & $\times 10^{-5}$ & $\times 10^{-4}$ }
\rotate \startdata
 6 Nov. 1998&$4.7 \pm 0.4$&$171 \pm 16$& $3 \times 10^{-3}$&6.5 $\pm$ 4.3
        &$<$ 5.0 &$<$ 1 \\
 7 Nov. 1998&$3.7 \pm 0.3$&$24 \pm 2$&$6 \times 10^{-4}$&33 $\pm$ 0.6
        &3.4 $\pm$ 6.1&$<$  2 \\
 8 Nov. 1998&$6.8 \pm 1.4$&$18 \pm 2$&$5 \times 10^{-4}$&23 $\pm$ 10
        &5.3 $\pm$ 11.1&$<$ 14 \\
 14 Nov. 1998&$1.78 \pm 0.05$&$21.9 \pm 0.5$&$5 \times 10^{-4}$&1.1 $\pm$ 0.3 \tablenotemark{\diamondsuit}
         &1.7 $\pm$ 2.3&$<$ 50 \\
 22 Nov. 1998&$1.9 \pm 0.4$&$22 \pm 4$&$5 \times 10^{-4}$&$<$ 2.8
         &5.1 $\pm$ 9.0&$<$ 9 \\
 24 Nov. 1998&$3.5 \pm 0.2$&$17 \pm 1$&$5 \times 10^{-4}$&4.1 $\pm$ 3.2
        &35 $\pm$ 1.4&$<$ 7 \\
 20 Jan. 1999&$2.8 \pm 0.2$&$182 \pm 14$&$3 \times 10^{-3}$&0.3 $\pm$ 0.6
        &3.5 $\pm$ 2.8&$<$ 6\\
 22 Jan. 1999&$4.2 \pm 0.1$&$166 \pm 8$&$3 \times 10^{-3}$&-0.1 $\pm$ 0.6
        &0.7 $\pm$ 1.1&$<$  3 \\
 16 Feb. 1999&$3.4 \pm 0.7$&$12.2 \pm 2.5$&$4 \times 10^{-4}$&-0.1 $\pm$ 8.0
        &37 $\pm$ 80&$<$ 90 \\

\enddata
\tablenotetext{\S}{Energy: 10--50 MeV n$^{-1}$}
\tablenotetext{\flat}{Energy: 12--14 MeV n$^{-1}$}
\tablenotetext{\dag}{$^{3}$He =$>$ $^{3}$He$_{SEP}$ -
$^{3}$He$_{BG}$; $\,\,\,\,$Energy: 15--45 MeV n$^{-1}$}
\tablenotetext{\sharp}{Energy: 9--12 MeV n$^{-1}$}
\tablenotetext{\ddag}{Energy: 9--12 MeV n$^{-1}$}
 \tablenotetext{\diamondsuit}{Energy $>$ 25 MeV
n$^{-1}$.}
\end{deluxetable}
\clearpage

\end{document}